\documentclass[preprint,showpacs,preprintnumbers,amsmath,amssymb]{revtex4}

\usepackage{graphicx}
\usepackage{dcolumn}
\usepackage{bm}

\begin{document}


\title{%
Topological properties of spin-triplet superconductors and the Fermi
surface topology in the normal state
}

\author{Masatoshi Sato}
\email{msato@issp.u-tokyo.ac.jp}
\affiliation{%
The Institute for Solid State Physics, The University of Tokyo,\\
Kashiwanoha 5-1-5, Kashiwa, Chiba, 277-8581, Japan,
}%


\date{\today}

\begin{abstract}
We report intimate relations between topological
 properties of full gapped spin-triplet superconductors with
 time-reversal invariance and the Fermi surface topology in the normal states.
An efficient method to calculate the ${\bm Z}_2$ invariants and the
 winding number for the spin-triplet superconductors is developed, and connections between these topological
 invariants and the Fermi surface structures in the normal states are
 pointed out.
We also obtain a correspondence between the Fermi
 surface topology and gapless surface states in the superconducting states.
The correspondence is inherent to spin-triplet superconductivity.
\end{abstract}

\pacs{
}
\maketitle

Search for possible states of quantum matter is one of the central
issues in condensed matter physics.
In addition to local order parameters,
gapped states can be characterized by topological invariants which are
constructed from the wave functions.
The quantum Hall state is a prominent example of such topological states,
in which the Hall conductance is identified with the topological number
introduced by Thouless, Kohmoto, Nightingale and den Nijs (TKNN)
\cite{TKNN82}.
While the TKNN number is non-trivial only when the time-reversal
symmetry is absent, recently new topological invariants, namely the
${\bm Z}_2$ invariants, were introduced in order to distinguish
a topological state with time-reversal invariance from ordinary band
insulators \cite{KM05a,MB07,Roy06,
FK06, FK07, FH07, SWSH06}.
For such a``topological insulator', 
the bulk-edge correspondence between the topological invariants in the
bulk and gapless edge (or
surface) states on the boundary was discussed 
in a similar manner to the quantum Hall state \cite{
KM05a, FK06, QWZ06}.
The topologically protected gapless state is an origin of
dissipationless (spin) Hall effects, which inspire an application to
spintronics \cite{MNZ04, BZ06}.  

In this paper, using the topological invariants, we study topological
properties of another class of gapped systems, spin-triplet superconductors.
Although conventional $s$-wave superconductors are topologically trivial,
it is known that an unconventional superconductor can be topologically
non-trivial \cite{Volvik01, HRK04, Sato06}.
In the following, we will develop a powerful method to evaluate the topological invariants for
spin-triplet superconductors, and find an intimate relation between
the topological properties in the spin-triplet superconducting state and those in the
normal state.
In particular, from topological arguments based on the bulk-edge correspondence, 
we derive formulas between the gapless surface (edge) state on the
boundary of the three dimensional (3D) (two dimensional (2D)) spin-triplet
superconductor and the topological invariants of the Fermi surface in
the normal state.
Although the number $N_0$ of the gapless surface (or edge) states
itself depends on the details of the gap function,
it will be shown that the index $(-1)^{N_0}$ does not depend on them
and is directly related to the Fermi
surface topology in the normal state. 
We also introduce a tight-binding lattice model, and confirm the
results by numerical calculations.   

In the following, we consider mainly full gapped spin-triplet superconductors with time-reversal
invariance.
A generalization to those without time-reversal invariance will be
mentioned in the last part of this paper briefly.

Let us start with the single-band description of
a spin-triplet superconducting state with time-reversal invariance.
(Generalization to the multi-band description is presented later.)
The Hamiltonian ${\cal H}$ of a spin-triplet superconductor in a
single-band  is given by 
\begin{eqnarray}
{\cal H}=\sum_{{\bm k},\sigma}\epsilon({\bm k})
c_{{\bm  k},\sigma}^{\dagger}c_{{\bm k},\sigma}+\frac{1}{2}
\sum_{{\bm k},\sigma,\sigma'}\left(
 \Delta_{\sigma \sigma'}({\bm k})
c_{{\bm k},\sigma}^{\dagger}c_{-{\bm k},\sigma'}^{\dagger} 
+{\rm h.c}\right),
\end{eqnarray}
where $c_{{\bm k},\sigma}^{\dagger}$ ($c_{{\bm k},\sigma}$) denotes a
creation (annihilation) operator of the electron, 
$\epsilon(\bm k)$ the dispersion of the electron in the normal
state and $\Delta(\bm k)$ the gap function given by
$
\Delta({\bm k})=i{\bm d}({\bm k})\cdot{\bm \sigma}\sigma_2.
$
${\bm d}({\bm k})$ are odd functions and ${\bm \sigma}$ are
the Pauli matrices.
By rewritten ${\cal H}$ as 
\begin{eqnarray}
{\cal H}=\frac{1}{2}\sum_{{\bm k}}{\bm c}_{\bm k}^{\dagger}H({\bm
 k}) {\bm c}_{{\bm k}},
\quad
{\bm c}_{\bm k}^{\dagger}=(c_{{\bm k},\sigma}^{\dagger}, c_{-{\bm k},\sigma}),
\end{eqnarray}
it is found that the spin-triplet superconducting state is
describe by the $4\times 4$ Bogoliubov-de Gennes (BdG) Hamiltonian,
\begin{eqnarray}
H({\bm k})=
\left(
\begin{array}{cc}
\epsilon({\bm k}){\bm 1}_{2\times 2} & \Delta({\bm k})\\
\Delta({\bm k})^{\dagger} & -\epsilon({\bm k}){\bm 1}_{2\times 2}
\end{array}
\right). 
\end{eqnarray}
We assume that the normal state has the inversion symmetry and the
time-reversal invariance, so $\epsilon(-{\bm k})=\epsilon({\bm k})$.
From the time-reversal invariance of $H({\bm k})$
\begin{eqnarray}
\Theta H({\bm k})\Theta^{-1}=H(-{\bm k})^{*},
\quad
\Theta=
\left(
\begin{array}{cc}
i\sigma_2 &0 \\
0 &i\sigma_2
\end{array}
\right),
\label{eq:theta}
\end{eqnarray}
${\bm d}({\bm k})$ should be real.
Eigenstates of $H({\bm k})$ with negative energies $E({\bm k})<0$ are occupied in the
ground state of the superconducting state.

The essential ingredient of our argument is the following ``symmetry'' for
the spin-triplet superconductor.
Since the parity of the gap function is odd, ${\bm d}(-{\bm k})=-{\bm
d}({\bm k})$,
the BdG Hamiltonian of the spin-triplet superconductor has the symmetry
\begin{eqnarray}
\Pi H({\bm k}) \Pi^{\dagger}=H(-{\bm k}),  
\quad \Pi^2=1
\label{eq:pi}
\end{eqnarray}
with 
\begin{eqnarray}
\Pi=\left(
\begin{array}{cc}
{\bm 1}_{2\times 2} &0 \\
0 & -{\bm 1}_{2\times 2}
\end{array} 
\right)=
{\bm 1}_{2\times 2}\otimes \tau_3.
\end{eqnarray}
Using this, we will study topological properties for
the spin triplet superconductor.

Let us consider
special points ${\bm k}=\Gamma_a$ in the Brillouin zone which are
time-reversal invariant and  satisfy $-\Gamma_a=\Gamma_a+{\bm G}$ for a
reciprocal-lattice vector ${\bm G}$.
In terms of the primitive
reciprocal lattice vectors ${\bm b}_j$, 
the time-reversal invariant momenta $\Gamma_a$ are
expressed as
\begin{eqnarray}
&&\Gamma_{a=(n_1,n_2)}=(n_1{\bm b}_1+n_2{\bm b}_2)/2
\quad
\mbox{for two dimensions},
\\ 
&&\Gamma_{a=(n_1,n_2,n_3)}=(n_1{\bm b}_1+n_2{\bm b}_2+n_3{\bm b}_3)/2, 
\quad
\mbox{for three dimensions},
\end{eqnarray}
with $n_j=0,1$.
At these momenta, 
the time-reversal invariance (\ref{eq:theta}) reduces to
$
\Theta H(\Gamma_a) \Theta^{-1}=H(\Gamma_a)^{*},
$
since $H({\bm k})$ satisfies $H({\bm k}+{\bm G})=H({\bm
k})$.
This implies that an occupied eigenstate $|u_n(\Gamma_a)\rangle$
$(n=1,2)$ has the same energy as its Kramers partner
$\Theta|u_n(\Gamma_a)\rangle^*$.
In addition, from the additional symmetry (\ref{eq:pi}),
we have
$
[H(\Gamma_a),\Pi]=0. 
$
So the Kramers doublet of the occupied states has the same eigenvalue of $\Pi$.
The eigenvalue of $\Pi$ is given by
\begin{eqnarray}
\pi_a=-{\rm sgn}\epsilon(\Gamma_a)
\label{eq:z2single}
\end{eqnarray}
since $H(\Gamma_a)=\epsilon(\Gamma_a)\Pi$.

The eigenvalues $\{\pi_a\}$ have
the following interesting properties: a) They are defined only at the
time-reversal momenta $\{\Gamma_a\}$. b) They only take $\pi_a=\pm 1$. c) Their
values can change only when the gap of the system closes.
To see the last property c), consider the quasiparticle spectrum,
$E({\bm k})=\pm \sqrt{\epsilon({\bm k})^2+{\bm d}({\bm k})^2}$,
which is obtained by diagonalizing $H({\bm k})$.
The gap of the system $2|E({\bm k})|$ closes when $\epsilon({\bm
k})={\bm d}({\bm k})=0$.
At the time-reversal momenta, the ${\bm d}$ vector vanishes
identically, ${\bm d}(\Gamma_a)=0$, so only $\epsilon(\Gamma_a)=0$ is
required for gap closing. 
Therefore, the gap closes when $\pi_a$ changes. 

The above properties suggest a connection between
the ${\bm Z}_2$ invariants introduced in \cite{KM05a} and $\{\pi_a\}$: 
The ${\bm Z}_2$ numbers are calculated from the
quantities $\{\delta_a\}$ \cite{FK07}
\begin{eqnarray}
\delta_a=\frac{\sqrt{{\rm det}[w(\Gamma_a)]}}{{\rm Pf}[w(\Gamma_a)]},
\label{eq:delta} 
\end{eqnarray} 
where $w(\Gamma_a)_{nm}$ is the anti-symmetric U(2) matrix
connecting the occupied states $|u_n(\Gamma_a)\rangle $ $(n=1,2)$ with
their Kramers partners 
$\Theta |u_n(\Gamma_a)\rangle^* $,
$w(\Gamma_a)_{nm}\equiv\langle u_n(\Gamma_a)|\Theta
|u_m(\Gamma_a)\rangle^{*}$, and ${\rm Pf}$ denotes its Pfaffian.
While the quantities $\{\delta_a\}$ depend on the gauge (or phase choice)
of the occupied states, their gauge-independent combinations define
the ${\bm Z}_2$ invariants, $\nu$ for two dimensions and $\nu_{\mu}$
($\mu=1,2,3,0$) for three dimensions \cite{FK07}:
$(-1)^{\nu}=\prod_{n_j=0,1}\delta_{a=(n_1,n_2)}$
for two dimensions, and 
$(-1)^{\nu_0}=\prod_{n_j=0,1}\delta_{a=(n_1,n_2,n_3)}$,
$(-1)^{\nu_k}=\prod_{n_{j\neq k}=0,1; n_k=1}\delta_{a=(n_1,n_2,n_3)}$ $(k=1,2,3)$
for three dimensions.
We notice here that the quantities $\{\delta_a\}$ have properties
similar to those of $\{\pi_a\}$:
A) They are defined only at the time-reversal invariant momenta.
B) They only take $\delta_a=\pm 1$ since ${\rm Pf}[w(\Gamma_a)]^2={\rm
det}[w(\Gamma_a)]$.
C) With fixing the gauge (or phase choice) of the occupied states, their
values can change only when the gap of the system closes. 
The last property C) is obvious because their gauge independent
combinations $\nu$ and $\nu_{\mu}$ can change only when the gap of the system closes.

These similarities suggest that the relation $\delta_a=\pi_a$
holds with a suitable phase choice of the occupied states.
Indeed, we can prove it by using a similar technique developed in
\cite{FK07} with the replacement of the inversion symmetry $P$ by the symmetry
$\Pi$ in the argument.
As a result, we obtain useful formulas of the ${\bm Z}_2$ invariants
for the time-reversal invariant spin-triplet superconductor,
\begin{eqnarray}
&&(-1)^{\nu}=\prod_{n_j=0,1}{\rm sgn}\epsilon(\Gamma_{a=(n_1,n_2)}),
\quad
\mbox{for two dimensions},
\label{eq:2dz2}
\\ 
&&
(-1)^{\nu_0}=\prod_{n_j=0,1}{\rm sgn}\epsilon(\Gamma_{a=(n_1,n_2,n_3)}),
\quad
(-1)^{\nu_k}=\prod_{n_{j\neq k}=0,1; n_k=1}
{\rm sgn}\epsilon(\Gamma_{a=(n_1,n_2,n_3)}),
\quad
\mbox{for three dimensions}.
\nonumber\\
\label{eq:3dz2}
\end{eqnarray}
Note that the ${\bm Z}_2$ numbers $\nu$ and $\nu_\mu$ are mod 2 integers,
which are identified
with $\nu+2$ and $\nu_{\mu}+2$, respectively.
Thus the ${\bm Z}_2$ numbers are non-trivial (trivial) when they are odd
(even).

\begin{figure}[h]
\begin{center}
\includegraphics[width=10cm]{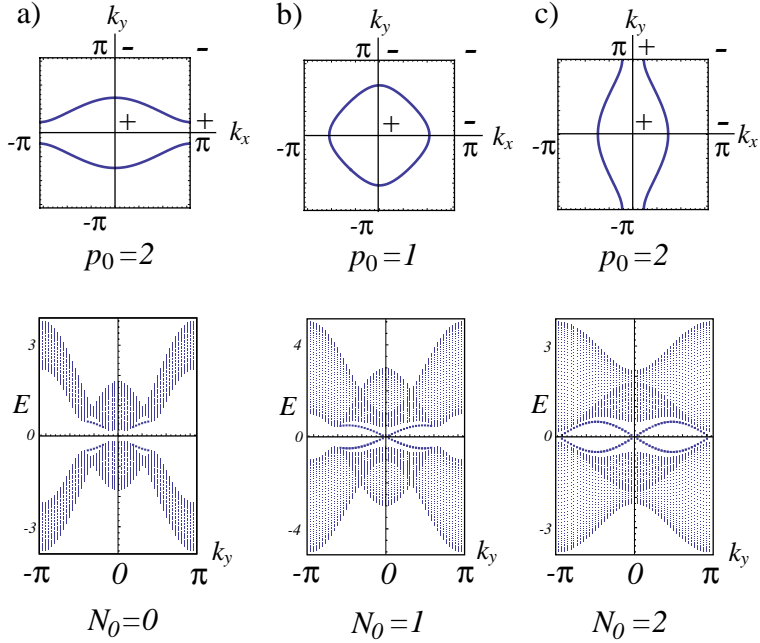}
\caption{(color online). The Fermi surfaces in the normal state and the edge states in
 2D time-reversal invariant spin-triplet superconducting state.
(Top row) The Fermi surfaces and $\pi_a=\pm$ at the
 time-reversal invariant momenta. $p_0(=p_0(S_{\rm F}))$ is the number of the
 connected components of the Fermi surface.
(Bottom row) The energy spectra of the corresponding superconducting
 states described by (\ref{eq:latticeH}) with edges at $i_x=0$ and
 $i_x=50$. Here $k_y$ denotes the momentum in the $y$-direction, and
 $N_0$ the number of gapless helical edge states. We set the parameters
 of the lattice model (\ref{eq:latticeH}) as (a) $t_x=0.4$, $t_y=1$,
 $\mu=-1$, $d=0.5$, (b) $t_x=t_y=1$, $\mu=-1$, $d=0.5$, and (c) $t_x=1$,
 $t_y=0.4$, $\mu=-1$, $d=0.5$, respectively.}
\label{fig:2dedgestate}
\end{center}
\end{figure}

\begin{figure}[h]
\begin{center}
\includegraphics[width=12cm]{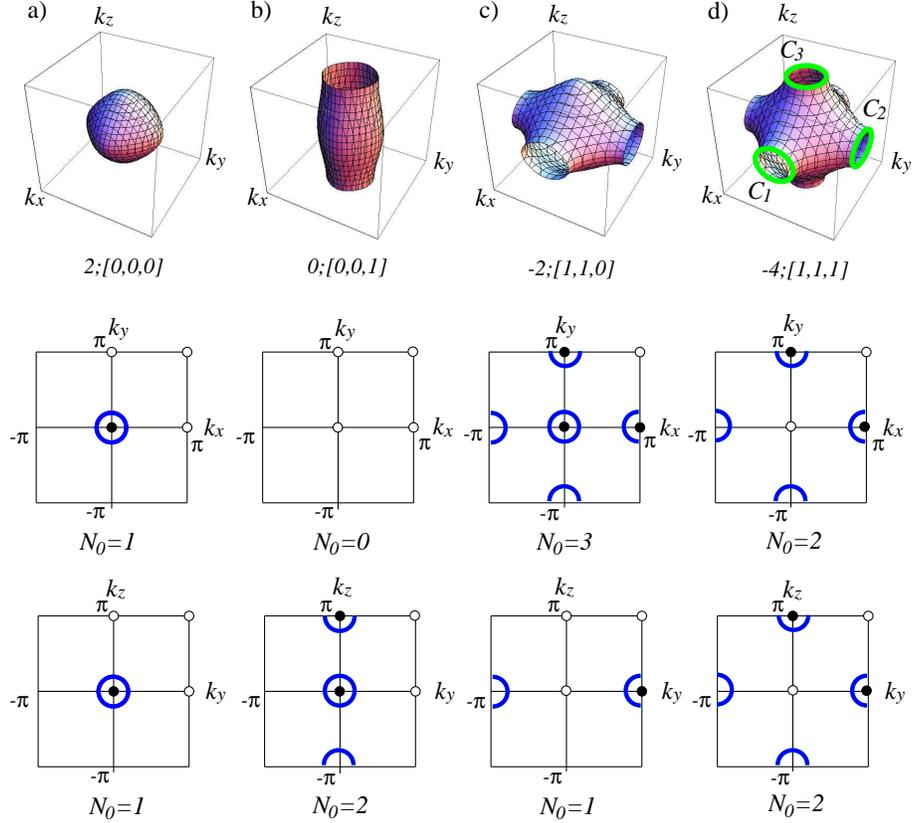}
\caption{(color online). 
Various Fermi surface topologies in three dimensions and the
 corresponding gapless surface states for
the 3D time-reversal invariant spin-triplet superconductor with ${\bm d}({\bm k})=\sin k_x \hat{\bm x}+
 \sin k_y \hat{\bm y}+\sin k_z \hat{\bm z}$.
(Top row) Fermi surfaces in the first Brillouin zone and their topological
 invariants, $\chi(S_F);[p_0(C_1),p_0(C_2),p_0(C_3)]$. The green circles
 are $C_j$ ($j=1,2,3$) for d). 
(Middle and bottom rows) The corresponding surface states on the Brillouin zone for 001
 surface (middle) and 100 surface (bottom) in the superconducting state.
The blue solid circles symbolize the Dirac cones of the 2D gapless
 Majorana fermions, and the energies of the surface states become zero at
 the time-reversal invariant momenta enclosed by the blue circles.  
 $N_0$ denotes the number of 2D gapless Majorana states on each
 surface.
}
\label{fig:3dtopology}
\end{center}
\end{figure}

Here we find that the right-hand sides of Eqs.(\ref{eq:2dz2}) and
(\ref{eq:3dz2}) have their
own topological meanings
related to the Fermi surface structure in the normal state:
For Eq.(\ref{eq:2dz2}), by using the relation $\epsilon(-{\bm k})=\epsilon({\bm
k})$, it is found that
\begin{eqnarray}
\prod_{n_j=0,1}{\rm sgn}\epsilon(\Gamma_{a=(n_1,n_2)})=(-1)^{p_0(S_{F})},
\label{eq:fermitopo1}
\end{eqnarray}
where $p_0(S_F)$ is the number of
different connected components of the Fermi surface in the normal state.
Also for Eq (\ref{eq:3dz2}), we obtain
\begin{eqnarray}
&&\prod_{n_j=0,1}{\rm
sgn}\epsilon(\Gamma_{a=(n_1,n_2,n_3)})=(-1)^{\chi(S_F)/2},
\nonumber\\
&&\prod_{n_{j\neq k}=0,1;n_k=1}{\rm sgn}\epsilon(\Gamma_{a=(n_1,n_2,n_3)})
=(-1)^{p_0(C_k)}, 
\label{eq:fermitopo2}
\end{eqnarray}
where $\chi(S_F)$ is the 
Euler characteristic of the Fermi surface, and $p_0(C_k)$ is the number of
different connected components of the intersection $C_k$ between the
Fermi surface 
and the time-reversal invariant plane with ${\bm k}={\bm b}_k/2$.
 (For a single
connected Fermi surface, the
Euler characteristic is given by $\chi(S_F)=2(1-g)$ with $g$ the genus of the
Fermi surface. When there are 
multiple connected components of the Fermi surface, $\chi(S_F)$  is the
sum of the
Euler characteristics of each component.)
We illustrate $p_0(S_{\rm F})$, $\chi(S_{\rm F})$ and
$p_0(C_k)$ in
Figs.\ref{fig:2dedgestate} and \ref{fig:3dtopology}.
Eqs.(\ref{eq:fermitopo1}) and (\ref{eq:fermitopo2}) are confirmed by
these examples.
These quantities, $p_0(S_F)$, $\chi(S_F)$ and $p_0(C_k)$ 
are topological invariants of the Fermi surface, and they do not change
the values under deformations of the Fermi surface unless the Fermi
surface crosses one of
the time-reversal invariant momenta. 
Therefore, Eqs.(\ref{eq:2dz2})-(\ref{eq:fermitopo2}) make connections
between the topological invariants in two different phases, {\it i.e,.}
the ${\bm Z}_2$ invariants in the superconducting phase and $p_0(S_F)$,
$\chi(S_F)$ and $p_0(C_k)$ in the normal phase:
\begin{eqnarray}
&& (-1)^{\nu}=(-1)^{p_0(S_{\rm F})},
\quad 
\mbox{for two dimensions},
\label{eq:2d}
\\ 
&& (-1)^{\nu_0}=(-1)^{\chi(S_{\rm F})/2}, \quad
(-1)^{\nu_k}=(-1)^{p_0(C_k)},
\quad 
\mbox{for three dimensions},
\label{eq:3d}
\end{eqnarray}

An important physical consequence of our formulas (\ref{eq:2d}) and
(\ref{eq:3d}) is that one can obtain useful information about gapless surface
(or edge) states in the spin-triplet superconductor from the knowledge
of the Fermi surface topology.
From the bulk-edge correspondence, a non-trivial ${\bm Z}_2$ number of a bulk gapped
system implies the existence of a gapless state localized on the
boundary \cite{FK06}.
For time-reversal invariant systems in two dimensions, the gapless
state is non-chiral and its Kramers doublet forms a helical pair \cite{BZ06,QHRZ08}.
The helical edge pair also satisfies the Majorana
condition in the present case, because of the particle-hole symmetry of the
superconducting system.
From a topological argument similar to that in \cite{FK06}, it is shown
that an odd (even) number of gapless helical Majorana pairs exist on each
edge when $(-1)^{\nu}=-1$ $((-1)^{\nu}=1)$.
Thus from (\ref{eq:2d}), we find the following connection between the
number $N_0$ of the gapless helical Majorana pairs on each edge and the
topological invariant $p_0(S_F)$ of the Fermi surface, 
\begin{eqnarray}
(-1)^{N_0}=(-1)^{p_0(S_F)}. 
\label{eq:bulkedge2d}
\end{eqnarray}
This formula implies that when $p_0(S_F)$ is odd, $N_0$ cannot be zero, and at
least one gapless helical Majorana state should exist on each edge.

For 3D time-reversal invariant spin-triplet superconductors, the
gapless boundary state
is a 2D massless Majorana fermion.
By generalizing the argument in \cite{FK06} to this case,
we have the following two properties of the surface state. 
1) The number $N_0$ of 2D gapless Majorana fermions
on a boundary surface is related to the topological number $\nu_0$ by
the equation $(-1)^{N_0}=(-1)^{\nu_0}$.
2) When $(-1)^{\nu_0}=1$, 
a non-trivial $\nu_i$ implies the existence of  2D gapless Majorana
fermions on surfaces determined by $\nu_i$:
To specify the surfaces,
consider a surface ${\bm G}$
which is perpendicular to a reciprocal-lattice vector ${\bm G}$.
If the surface ${\bm G}$ satisfies ${\bm G}\neq \sum_i
(\nu_i+2m_{i}){\bm b}_i$ for any integers $m_i$, 
then there exist 2D gapless Majorana fermions on the surface.
Combining the former property with (\ref{eq:3d}), we have a relation
between the gapless surface state of a 3D time-reversal invariant
spin-triplet superconductor and its Fermi surface topology as,
\begin{eqnarray}
(-1)^{N_0}=(-1)^{\chi(S_F)/2},
\label{eq:bulkedge3d}
\end{eqnarray}
where $N_0$ the number of the 2D gapless Majorana fermions on the boundary
surface. 
Moreover, taking into account the latter property as well, we
obtain the following predictions.
(i) When the Fermi surface satisfies $(-1)^{\chi(S_F)/2}=-1$,  an
odd number of 2D gapless 
Majorana fermions exist on each boundary surface.
In particular, at least one gapless Majorana fermion exists on each
boundary surface.
(ii) When the Fermi surface satisfies  $(-1)^{\chi(S_F)/2}=1$,
the number of the 2D gapless Majorana fermions on a boundary surface
is even.
Then if the surface ${\bm G}$ satisfies  ${\bm G}\neq \sum_i (p_0(C_i)+2m_i){\bm
b}_i$ with arbitrary integers $m_i$,
at least two 2D massless Majorana fermions exist on the boundary surface
${\bm G}$.
On the other hand, if ${\bm G}=\sum_i(p_0(C_i)+2m_i){\bm b}_i$ with
integers $m_i$, no gapless Majorana fermion is possible on the surface
${\bm G}$.

In Table \ref{table}, 
we summarize the relations between the
Fermi surface topology and the boundary gapless state \footnote{Note that in
order to determine which
$N_0$ in Table I is realized in a time-reversal invariant spin-triplet
superconductor, we need the knowledge of the gap function in addition to
that of the Fermi surface structure. Nevertheless, for the cases in the first
row in Table \ref{table} a) and the first and third rows in Table
\ref{table} b), we can conclude that $N_0$ becomes non-zero only from the
knowledge of the Fermi surface topology.}. 
Later, we will check these results by using concrete models.

\begin{table}

\begin{flushleft}
\hspace{3ex} a) 2D case
\end{flushleft}
 
\begin{tabular}{|l||c|}
\hline
$(-1)^{p_0(S_{\rm F})}=-1$ & $N_0=1,3,5, \cdots $
\\ \hline
$(-1)^{p_0(S_{\rm F})}=1$ & $N_0=0,2,4, \cdots $
\\ \hline
\end{tabular}
 
\begin{flushleft}
\hspace{3ex} b) 3D case
\end{flushleft}
\begin{tabular}{|c||c|c|}
\hline
$(-1)^{\chi(S_{\rm F})/2}=-1$ & \multicolumn{2}{c|} {$N_0=1,3,5, \cdots $}
\\ \hline
&
\mbox{On a surface ${\bm G}=\sum_i (p_0(C_i)+2 m_i){\bm b}_i$}
& $N_0=0,2,4,\cdots $
\\ \cline{2-3}
\raisebox{2.5ex}[0pt]{$(-1)^{\chi(S_{\rm F})/2}=1$}
& \mbox{On a surface ${\bm G}\neq \sum_i (p_0(C_i)+2
     m_i){\bm b}_i$} & $N_0=2,4,6,\cdots $
\\ \hline
 \end{tabular}
 
\caption{Topological invariants of the Fermi surface and the possible
 number $N_0$ of gapless boundary states for
 full gapped time-reversal invariant spin-triplet superconductors. a) 2D
 case. Here $p_0(S_{\rm F})$ denotes the number of connected components of
 the Fermi surface, and $N_0$ the possible number of gapless helical Majorana
 pairs on an edge. b) 3D case. Here $\chi(S_{\rm F})$ is the Euler
 characteristic of the Fermi surface, $p_0(C_k)$ the number of different
 connected components of the intersection $C_k$ between the Fermi
 surface and the time-reversal invariant plane with ${\bm k}={\bm
 b}_k/2$, and $m_i$ integers. The surface ${\bm G}$ is perpendicular to
 the reciprocal-lattice
 vector ${\bm G}$, and $N_0$ is the possible number of 2D Majorana fermion
 on the surface ${\bm G}$.}
\label{table}
\end{table}

For 3D
time-reversal invariant spin-triplet superconductors, it is also known
that there exists
another topological invariant $\nu_{\rm w}$ called the winding number
\cite{GV88, SRFL08}.
Now we will derive a useful formula for $\nu_{\rm w}$ and show that
$\nu_{\rm w}$ also has an
intimate relation to the Fermi surface topology. 
In the single band description, the winding number $\nu_{\rm w}$ is
given by
\begin{eqnarray}
\nu_{\rm w}=\frac{1}{12\pi^2}\int_{T^3}dk^3\epsilon^{ijk}\epsilon^{abcd}
\hat{\eta}_a 
\partial_i\hat{\eta}_b \partial_j\hat{\eta}_c \partial_k\hat{\eta}_d,
\label{eq:winding3}
\end{eqnarray}
where $T^3$ denotes the first Brillouin zone, and
$\hat{\eta}_a({\bm k})=\eta_a({\bm k})/\sqrt{\eta_a({\bm k})^2}$ with
$\eta_a({\bm k})=({\bm d}({\bm k}),\epsilon({\bm k}))$.
$\nu_{\rm w}$ counts the number of times the unit vector $\hat{\eta}_a$
wraps the 3D sphere ${\rm S}^3$ ($\hat{
\eta}_a^2=1$) when we sweep $T^3$.
In order for $\hat{\eta}_a$ to wind ${\rm S}^3$, it is necessary to pass
the poles of ${\rm S}^3$ defined by ${\bm
\eta}\equiv(\eta_1,\eta_2,\eta_3)\equiv {\bm d}={\bm 0}$. 
So consider the set of zeros ${\bm k}^*$ satisfying ${\bm
\eta}({\bm k}^{*})={\bm 0}$.
From the topological nature of $\nu_{\rm w}$, we can rescale
$\epsilon({\bm k})$ as $\epsilon({\bm k})\rightarrow a\epsilon({\bm k})$
$(a\ll 1)$ without changing the value of $\nu_{\rm w}$.
Then it is found that only neighborhoods of the zeros contribute to
$\nu_{\rm w}$ if $a$ is small enough.
By expanding $\eta_a$ as
$
\eta_i=\partial_j d_i({\bm k}^*)({\bm k}-{\bm k}^{*})_j+\cdots,
\, (i=1,2,3)
$,
$
\eta_4=\epsilon({\bm k}^*)\, (\ll 1)
$,
the contribution from the zero ${\bm k}^{*}$ is evaluated as
\begin{eqnarray}
\nu_{\rm w}({\bm k}^*)=-\frac{1}{2}{\rm sgn}(\epsilon({\bm k}^{*})){\rm sgn}
\left({\rm det}(\partial_j d_i({\bm k}^*))\right). 
\label{eq:windinglocal0}
\end{eqnarray}
(When ${\rm det}(\partial_j d_i({\bm k}^*))=0$,
Eq.(\ref{eq:windinglocal0}) is generalized to
\begin{eqnarray}
\nu_{\rm w}({\bm k}^*)=-\frac{1}{2}{\rm sgn}(\epsilon({\bm k}))i({\bm k}^*),
\label{eq:windinglocal}
\end{eqnarray}
where $i({\bm k}^*)$ denotes the Poincar\'{e}-Hopf index \cite{AFG75} of
the zero ${\bm k}^{*}$.)
Summing up the contributions of all zeros, we have
\begin{eqnarray}
\nu_{\rm w}=\sum_{{\bm \eta}({\bm k}^{*})={\bm 0}}\nu_{\rm w}({\bm k}^{*}).
\label{eq:winding4}
\end{eqnarray}
From (\ref{eq:winding4}), we can show that $\nu_{\rm w}$ is also related
to $\chi(S_F)$. 
For simplicity, suppose that the set of zeros ${\bm k}^{*}$ contains only the
time-reversal invariant points $\{\Gamma_a\}$.
($\Gamma_a$ is always a zero since it satisfies ${\bm
d}(\Gamma_a)={\bm 0}$.)
Dividing the set of zeros into two subsets, $\Gamma_{\pm}\equiv \{\Gamma_a;
{\rm sgn}\epsilon(\Gamma_a)=\pm 1\}$, we obtain $\nu_{\rm
w}=-\sum_{\Gamma_a\in \Gamma_+}i(\Gamma_a)/2+\sum_{\Gamma_a\in \Gamma_-}i(\Gamma_a)/2$.
Then by using the Poincar\'{e}-Hopf theorem $\sum_{{\bm k}^{*}}i({\bm
k}^*)=0$ \cite{Milnor65},
it is recast into $\nu_{\rm w}=\sum_{\Gamma_a\in
\Gamma_-}i(\Gamma_a)$.
Here $i(\Gamma_a)$ is an odd integer because of ${\bm d}(-{\bm k})=-{\bm
d}({\bm k})$.
Therefore, $\nu_{\rm w}$ is an odd (even) integer if $\Gamma_{-}$ has an
odd (even) number of elements.
From this, we obtain the relation
\begin{eqnarray}
(-1)^{\nu_{\rm w}}=\prod_{n_j=0,1}{\rm
sgn}\epsilon(\Gamma_{a=(n_1,n_2,n_3)}).
\end{eqnarray}
Combining this with (\ref{eq:fermitopo2}) and (\ref{eq:bulkedge3d}), we
find that $\nu_{\rm w}$
is also related to the Euler characteristic $\chi(S_{\rm F})$ and the
number $N_0$ of 2D gapless surface states as
\begin{eqnarray}
(-1)^{\nu_{\rm w}}=(-1)^{\chi(S_{\rm F})/2}=(-1)^{N_0}. 
\label{eq:windingN0}
\end{eqnarray}

Let us now illustrate our results with simple and important examples.
In Fig.\ref{fig:2dedgestate}, we illustrate possible Fermi surfaces in
the normal state and the corresponding $p_0(S_F)$ in two dimensions.
We also present the energy spectra for the corresponding superconducting
states with edges.
To obtain the energy spectra,  we use the lattice model of
the superconducting state
with ${\bm d}=d(\sin k_x \hat{\bm x}+\sin k_y \hat{\bm y})$,
\begin{eqnarray}
{\cal H}=\frac{1}{2}\sum_{{\bm i}{\bm j}}{\bm c}^{\dagger}_{\bm i}
H_{{\bm i}{\bm j}}{\bm c}_{\bm j},
\quad
H_{{\bm i}{\bm j}}=
\left(
\begin{array}{cc}
t_{{\bm i}{\bm j}} & i{\bm d}_{{\bm i}{\bm j}}\cdot{\bm \sigma}\sigma_2\\
-i{\bm \sigma}\sigma_2{\bm d}_{{\bm j}{\bm i}} & -t_{{\bm i}{\bm j}}
\end{array} 
\right),
\label{eq:latticeH}
\end{eqnarray}
where ${\bm c}^{\dagger}_{\bm i}=(c_{{\bm i}\sigma}^{\dagger},c_{{\bm
 i}\sigma})$, and $t_{{\bm i}{\bm j}}$ and ${\bm d}_{{\bm i}{\bm j}}$
 are given by
$
t_{{\bm i}{\bm j}}=
-t_x(\delta_{{\bm i},{\bm j}+\hat{\bm x}}
+\delta_{{\bm j},{\bm i}+\hat{\bm x}})
-t_y(\delta_{{\bm i},{\bm j}+\hat{\bm y}}
+\delta_{{\bm j},{\bm i}+\hat{\bm y}})
-\mu\delta_{{\bm i}{\bm j}},
$
$
(d_x)_{{\bm i}{\bm j}}=-i(d/2)
(\delta_{{\bm j},{\bm i}+\hat{\bm x}}
-\delta_{{\bm i},{\bm j}+\hat{\bm x}}),
$
$
(d_y)_{{\bm i}{\bm j}}=-i(d/2)
(\delta_{{\bm j},{\bm i}+\hat{\bm y}}
-\delta_{{\bm i},{\bm j}+\hat{\bm y}}),
$
$
(d_z)_{{\bm i}{\bm j}}=0.
$
The spectra are calculated for the system with two edges at $i_x=0, 50$
 under the periodic boundary condition in the $y$-direction.
In Fig.\ref{fig:2dedgestate}, $k_y$ denotes the momentum in the $y$-direction. 
While no gapless edge state exists in Fig.\ref{fig:2dedgestate} a), 
it is found that there exist
gapless edge states in the bulk gap in Figs. \ref{fig:2dedgestate} b) and c).
The relation Eq.(\ref{eq:bulkedge2d}) holds in
 Fig.\ref{fig:2dedgestate}.

In Fig.\ref{fig:3dtopology}, we show various Fermi
surfaces in the first Brillouin zone and their topological numbers,
$\chi(S_F)$ and $p_0(C_i)$ $(i=1,2,3)$ in three dimensions.
In addition, we present gapless 2D Majorana surface states for the 
the superconducting states with
$
{\bm d}({\bm k})=\sin k_x\hat{\bm x}
+\sin k_y\hat{\bm y}+\sin k_z\hat{\bm z}. 
$
This figure also confirms the connection between the gapless surface
states and the Fermi surface topology:
The relation $(-1)^{\chi(S_{\rm F})/2}=(-1)^{N_0}$ holds for all the cases.
Furthermore, in the cases with $(-1)^{\chi(S_{\rm F})/2}=1$
({\it i.e.} Fig.\ref{fig:3dtopology} b) and d) ), there exist a non-zero even
number of 2D gapless Majorana fermions on a surface ${\bm G}\neq
\sum_i(p_0(C_i)+2m_i {\bm b}_i)$ with integers $m_i$.
Note that in Fig.\ref{fig:3dtopology}, only the 001 surface in
Fig.\ref{fig:3dtopology} b) does not satisfy this condition. In this
case, we have ${\bm G}={\bm b}_3$, and it coincides with
$\sum_i p_0(C_i) {\bm b}_i={\bm b}_3$.  
From (\ref{eq:winding4}), we find that $\nu_{\rm w}$'s for this gap
function are (a) $\nu_{\rm
w}=1$, (b) $\nu_{\rm w}=0$, (c) $\nu_{\rm w}=-1$, and (d) $\nu_{\rm
w}=-2$, respectively.
These values are also consistent with Eq.(\ref{eq:windingN0}).

So far we have considered the single-band superconductor. However,
the formulas (\ref{eq:2dz2}) and (\ref{eq:3dz2}) 
can be generalized to multi-band systems.
To see this, consider a multi-band system which has the inversion
symmetry and the time-reversal invariance in the normal state. 
If we assume that the parity operator transforms
only the momentum as ${\bm k}\rightarrow -{\bm k}$ \footnote{This
assumption is met for most cases, however, for a system with a sublattice
structure, the parity operator may also exchange the sublattice
structure (for example, see \cite{FK07}.)},
then the Hamiltonian in the normal state is given by a $2N\times 2N$ matrix
${\cal E}({\bm k})$ satisfying ${\cal E}(-{\bm k})={\cal E}({\bm k})$.
($N$ is the number of the bands.)
Odd-parity superconducting states for this system are described by the
generalized BdG Hamiltonian
\begin{eqnarray}
H({\bm k})=
\left(
\begin{array}{cc}
{\cal E}({\bm k}) & \Delta({\bm k})\\
\Delta({\bm k})^{\dagger} & -{\cal E}({\bm k})
\end{array} 
\right),
\end{eqnarray}
where the gap function $\Delta({\bm k})$ is a $2N\times 2N$ matrix with
odd parity, $\Delta(-{\bm k})=-\Delta({\bm k})$. 
$H({\bm k})$ has the property
\begin{eqnarray}
\Pi H({\bm k})\Pi^{\dagger} 
= H(-{\bm k}),
\quad
\Pi^2=1 
\end{eqnarray}
with 
$
\Pi={\bm 1}_{2N\times 2N}\otimes \tau_3,  
$
and for ${\bm k}=\Gamma_a$, $H({\bm k})$ becomes
$
H(\Gamma_a)={\cal E}(\Gamma_a)\otimes \tau_3,
$
thus in a similar manner as the single-band case, 
it is shown that
\begin{eqnarray}
&&(-1)^{\nu}=\prod_{n_j=0,1} 
\prod_{m=1}^{N}{\rm sgn}(E_{2m}(\Gamma_{a=(n_1,n_2)})),
\quad
\mbox{for two dimensions},
\label{eq:2dz2multi}
\\
&&(-1)^{\nu_0}=\prod_{n_j=0,1} 
\prod_{m=1}^{N}{\rm sgn}(E_{2m}(\Gamma_{a=(n_1,n_2,n_3)})),
\nonumber\\
&&(-1)^{\nu_k}=\prod_{n_{j\neq k}=0,1;n_k=1} 
\prod_{m=1}^{N}{\rm sgn}(E_{2m}(\Gamma_{a=(n_1,n_2,n_3)})),
\quad
\mbox{for three dimensions},
\label{eq:3dz2multi}
\end{eqnarray}
where $E_n(\Gamma_a)$ $(n=1,\cdots 2N)$ are the
eigenvalues of ${\cal E}({\bm k})$ at ${\bm k}=\Gamma_a$, and we have
set $E_{2m}(\Gamma_a)=E_{2m-1}(\Gamma_a)$ using the Kramers degeneracy.
For a filled or empty band in the normal state, the signatures of
$E_n(\Gamma_a)$ are the same for all the time-reversal points, so their
contributions to (\ref{eq:2dz2multi}) and (\ref{eq:3dz2multi}) are canceled.
Therefore, in order to evaluate the ${\bm Z}_2$ numbers, it is enough to
consider bands with the Fermi surfaces.  
Again it is evident that topological properties of the spin-triplet
superconducting state are closely related to the topology of the Fermi
surface.

Finally we make several comments in order.
a) Although we have assumed that
the normal state has the inversion symmetry, our formulas (\ref{eq:2dz2})
and (\ref{eq:3dz2}) (or (\ref{eq:2dz2multi}) and (\ref{eq:3dz2multi}))
could be useful
even for the systems which do not have the inversion symmetry in the normal
state: 
Adiabatic continuity allows us to calculate the
topological invariants if the system is adiabatically connected to
materials which have the inversion symmetry in the normal state.
The topological invariants for a class of noncentrosymmetric
superconductors can be calculated in this manner \cite{TYBN09,SF09}.
b) For spin-singlet superconductors, due to the inversion symmetry, their
${\bm Z}_2$ numbers are calculated by the technique developed in
\cite{FK07}. However, it is found that all the ${\bm Z}_2$ numbers are
trivial \cite{Sato09}. 
Therefore, the correspondence between the Fermi surface topology and
the gapless surface state discussed in this paper are inherent to
spin-triplet superconductors. 
c) In this paper we have focused on the time-reversal invariant spin-triplet
superconductors.
Here we mention a generalization to the time-reversal
breaking case in brief.  For 2D chiral spin-triplet
superconductors such as a $p+ip$ state,  the topological properties are
determined by the TKNN number $\nu_{\rm TKNN}$.
In a similar manner to $\nu_{\rm w}$, in the single-band description, it
can be shown that the TKNN number is related to the Fermi surface
topology by the equation 
\begin{eqnarray}
(-1)^{\nu_{\rm TKNN}}=(-1)^{p_0(S_F)}, 
\end{eqnarray}
where
$p_0(S_F)$ is the number of the connected components of the Fermi
surface \cite{Sato09}.
This relation gives a simple explanation of the quantum phase transition
from the weak paring phase to the
strong one discussed in \cite{RG00}.
This phase transition is accompanied with
disappearance of the Fermi surface, thus $p_0(S_F)=1\rightarrow p_0(S_F)=0$.
From the above relation, this cause a change of $\nu_{\rm TKNN}$, which
brings about different topological properties between the weak and
strong phases.

The author would like to thank Y. Asano, Y. Tanaka, X.-L. Qi, R. Roy, 
and Y.-S. Wu for helpful discussions.

\bibliography{topological_order}

\begin{thebibliography}{23}
\expandafter\ifx\csname natexlab\endcsname\relax\def\natexlab#1{#1}\fi
\expandafter\ifx\csname bibnamefont\endcsname\relax
  \def\bibnamefont#1{#1}\fi
\expandafter\ifx\csname bibfnamefont\endcsname\relax
  \def\bibfnamefont#1{#1}\fi
\expandafter\ifx\csname citenamefont\endcsname\relax
  \def\citenamefont#1{#1}\fi
\expandafter\ifx\csname url\endcsname\relax
  \def\url#1{\texttt{#1}}\fi
\expandafter\ifx\csname urlprefix\endcsname\relax\def\urlprefix{URL }\fi
\providecommand{\bibinfo}[2]{#2}
\providecommand{\eprint}[2][]{\url{#2}}

\bibitem[{\citenamefont{Thouless et~al.}(1982)\citenamefont{Thouless, Kohmoto,
  Nightingale, and den Nijs}}]{TKNN82}
\bibinfo{author}{\bibfnamefont{D.~J.} \bibnamefont{Thouless}},
  \bibinfo{author}{\bibfnamefont{M.}~\bibnamefont{Kohmoto}},
  \bibinfo{author}{\bibfnamefont{M.~P.} \bibnamefont{Nightingale}},
  \bibnamefont{and} \bibinfo{author}{\bibfnamefont{M.}~\bibnamefont{den Nijs}},
  \bibinfo{journal}{Phys.\ Rev.\ Lett.} \textbf{\bibinfo{volume}{49}},
  \bibinfo{pages}{405} (\bibinfo{year}{1982}).

\bibitem[{\citenamefont{Kane and Mele}(2005)}]{KM05a}
\bibinfo{author}{\bibfnamefont{C.~L.} \bibnamefont{Kane}} \bibnamefont{and}
  \bibinfo{author}{\bibfnamefont{E.~J.} \bibnamefont{Mele}},
  \bibinfo{journal}{Phys.\ Rev.\ Lett.} \textbf{\bibinfo{volume}{95}},
  \bibinfo{pages}{146802} (\bibinfo{year}{2005}).

\bibitem[{\citenamefont{Moore and Balents}(2007)}]{MB07}
\bibinfo{author}{\bibfnamefont{J.~E.} \bibnamefont{Moore}} \bibnamefont{and}
  \bibinfo{author}{\bibfnamefont{L.}~\bibnamefont{Balents}},
  \bibinfo{journal}{Phys.\ Rev.\ B} \textbf{\bibinfo{volume}{75}},
  \bibinfo{pages}{121306(R)} (\bibinfo{year}{2007}).

\bibitem[{\citenamefont{Roy}()}]{Roy06}
\bibinfo{author}{\bibfnamefont{R.}~\bibnamefont{Roy}},
  \eprint{arXiv:cond-mat/0608064}.

\bibitem[{\citenamefont{Fu and Kane}(2006)}]{FK06}
\bibinfo{author}{\bibfnamefont{L.}~\bibnamefont{Fu}} \bibnamefont{and}
  \bibinfo{author}{\bibfnamefont{C.~L.} \bibnamefont{Kane}},
  \bibinfo{journal}{Phys.\ Rev.\ B} \textbf{\bibinfo{volume}{74}},
  \bibinfo{pages}{195312} (\bibinfo{year}{2006}).

\bibitem[{\citenamefont{Fu and Kane}(2007)}]{FK07}
\bibinfo{author}{\bibfnamefont{L.}~\bibnamefont{Fu}} \bibnamefont{and}
  \bibinfo{author}{\bibfnamefont{C.~L.} \bibnamefont{Kane}},
  \bibinfo{journal}{Phys.\ Rev.\ B} \textbf{\bibinfo{volume}{76}},
  \bibinfo{pages}{045302} (\bibinfo{year}{2007}).

\bibitem[{\citenamefont{Fukui and Hatsugai}(2007)}]{FH07}
\bibinfo{author}{\bibfnamefont{T.}~\bibnamefont{Fukui}} \bibnamefont{and}
  \bibinfo{author}{\bibfnamefont{Y.}~\bibnamefont{Hatsugai}},
  \bibinfo{journal}{Phys.\ Rev.\ B} \textbf{\bibinfo{volume}{75}},
  \bibinfo{pages}{121403(R)} (\bibinfo{year}{2007}).

\bibitem[{\citenamefont{Sheng et~al.}(2006)\citenamefont{Sheng, Weng, Sheng,
  and Haldane}}]{SWSH06}
\bibinfo{author}{\bibfnamefont{D.~N.} \bibnamefont{Sheng}},
  \bibinfo{author}{\bibfnamefont{Z.~Y.} \bibnamefont{Weng}},
  \bibinfo{author}{\bibfnamefont{L.}~\bibnamefont{Sheng}}, \bibnamefont{and}
  \bibinfo{author}{\bibfnamefont{F.~D.} \bibnamefont{Haldane}},
  \bibinfo{journal}{Phys.\ Rev.\ Lett.} \textbf{\bibinfo{volume}{97}},
  \bibinfo{pages}{036808} (\bibinfo{year}{2006}).

\bibitem[{\citenamefont{Qi et~al.}(2006)\citenamefont{Qi, Wu, and
  Zhang}}]{QWZ06}
\bibinfo{author}{\bibfnamefont{X.-L.} \bibnamefont{Qi}},
  \bibinfo{author}{\bibfnamefont{Y.-S.} \bibnamefont{Wu}}, \bibnamefont{and}
  \bibinfo{author}{\bibfnamefont{S.-C.} \bibnamefont{Zhang}},
  \bibinfo{journal}{Phys.\ Rev.\ B} \textbf{\bibinfo{volume}{74}},
  \bibinfo{pages}{045125} (\bibinfo{year}{2006}).

\bibitem[{\citenamefont{Murakami et~al.}(2004)\citenamefont{Murakami, Nagaosa,
  and Zhang}}]{MNZ04}
\bibinfo{author}{\bibfnamefont{S.}~\bibnamefont{Murakami}},
  \bibinfo{author}{\bibfnamefont{N.}~\bibnamefont{Nagaosa}}, \bibnamefont{and}
  \bibinfo{author}{\bibfnamefont{S.~C.} \bibnamefont{Zhang}},
  \bibinfo{journal}{Phys.\ Rev.\ Lett.} \textbf{\bibinfo{volume}{93}},
  \bibinfo{pages}{156804} (\bibinfo{year}{2004}).

\bibitem[{\citenamefont{Bernevig and Zhang}(2006)}]{BZ06}
\bibinfo{author}{\bibfnamefont{B.~A.} \bibnamefont{Bernevig}} \bibnamefont{and}
  \bibinfo{author}{\bibfnamefont{S.~C.} \bibnamefont{Zhang}},
  \bibinfo{journal}{Phys.\ Rev.\ Lett.} \textbf{\bibinfo{volume}{96}},
  \bibinfo{pages}{106802} (\bibinfo{year}{2006}).

\bibitem[{\citenamefont{Volvik}(2001)}]{Volvik01}
\bibinfo{author}{\bibfnamefont{G.~E.} \bibnamefont{Volvik}},
  \bibinfo{journal}{Phys.\ Rep.} \textbf{\bibinfo{volume}{351}},
  \bibinfo{pages}{195} (\bibinfo{year}{2001}).

\bibitem[{\citenamefont{Hatsugai et~al.}(2004)\citenamefont{Hatsugai, Ryu, and
  Kohmoto}}]{HRK04}
\bibinfo{author}{\bibfnamefont{Y.}~\bibnamefont{Hatsugai}},
  \bibinfo{author}{\bibfnamefont{S.}~\bibnamefont{Ryu}}, \bibnamefont{and}
  \bibinfo{author}{\bibfnamefont{M.}~\bibnamefont{Kohmoto}},
  \bibinfo{journal}{Phys.\ Rev.\ B} \textbf{\bibinfo{volume}{70}},
  \bibinfo{pages}{054502} (\bibinfo{year}{2004}).

\bibitem[{\citenamefont{Sato}(2006)}]{Sato06}
\bibinfo{author}{\bibfnamefont{M.}~\bibnamefont{Sato}},
  \bibinfo{journal}{Phys.\ Rev.\ B} \textbf{\bibinfo{volume}{73}},
  \bibinfo{pages}{214502} (\bibinfo{year}{2006}).

\bibitem[{\citenamefont{Qi et~al.}(2009)\citenamefont{Qi, Hughes, Raghu, and
  Zhang}}]{QHRZ08}
\bibinfo{author}{\bibfnamefont{X.~L.} \bibnamefont{Qi}},
  \bibinfo{author}{\bibfnamefont{T.~L.} \bibnamefont{Hughes}},
  \bibinfo{author}{\bibfnamefont{S.}~\bibnamefont{Raghu}}, \bibnamefont{and}
  \bibinfo{author}{\bibfnamefont{S.~C.} \bibnamefont{Zhang}},
  \bibinfo{journal}{Phys.\ Rev.\ Lett.} \textbf{\bibinfo{volume}{102}},
  \bibinfo{pages}{187001} (\bibinfo{year}{2009}).

\bibitem[{\citenamefont{Grinevich and Volovik}(1988)}]{GV88}
\bibinfo{author}{\bibfnamefont{P.~G.} \bibnamefont{Grinevich}}
  \bibnamefont{and} \bibinfo{author}{\bibfnamefont{G.~E.}
  \bibnamefont{Volovik}}, \bibinfo{journal}{J.\ Low\ Temp.\ Phys.}
  \textbf{\bibinfo{volume}{72}}, \bibinfo{pages}{371} (\bibinfo{year}{1988}).

\bibitem[{\citenamefont{Schnyder et~al.}(2008)\citenamefont{Schnyder, Ryu,
  Furusaki, and Ludwig}}]{SRFL08}
\bibinfo{author}{\bibfnamefont{A.~P.} \bibnamefont{Schnyder}},
  \bibinfo{author}{\bibfnamefont{S.}~\bibnamefont{Ryu}},
  \bibinfo{author}{\bibfnamefont{A.}~\bibnamefont{Furusaki}}, \bibnamefont{and}
  \bibinfo{author}{\bibfnamefont{A.~W.~W.} \bibnamefont{Ludwig}},
  \bibinfo{journal}{Phys.\ Rev.\ B} \textbf{\bibinfo{volume}{78}},
  \bibinfo{pages}{195125} (\bibinfo{year}{2008}).

\bibitem[{\citenamefont{Arafune et~al.}(1975)\citenamefont{Arafune, Freund, and
  Goebel}}]{AFG75}
\bibinfo{author}{\bibfnamefont{J.}~\bibnamefont{Arafune}},
  \bibinfo{author}{\bibfnamefont{P.~G.~O.} \bibnamefont{Freund}},
  \bibnamefont{and} \bibinfo{author}{\bibfnamefont{C.~J.}
  \bibnamefont{Goebel}}, \bibinfo{journal}{J.\ Math.\ Phys.}
  \textbf{\bibinfo{volume}{16}}, \bibinfo{pages}{433} (\bibinfo{year}{1975}).

\bibitem[{\citenamefont{Milnor}(1965)}]{Milnor65}
\bibinfo{author}{\bibfnamefont{J.~W.} \bibnamefont{Milnor}},
  \emph{\bibinfo{title}{Topology from the Differential Viewpoint}}
  (\bibinfo{publisher}{Princeton University Press}, \bibinfo{year}{1965}).

\bibitem[{\citenamefont{Sato and Fujimoto}(2009)}]{SF09}
\bibinfo{author}{\bibfnamefont{M.}~\bibnamefont{Sato}} \bibnamefont{and}
  \bibinfo{author}{\bibfnamefont{S.}~\bibnamefont{Fujimoto}},
  \bibinfo{journal}{Phys.\ Rev.\ B} \textbf{\bibinfo{volume}{79}},
  \bibinfo{pages}{094504} (\bibinfo{year}{2009}).

\bibitem[{\citenamefont{Tanaka et~al.}(2009)\citenamefont{Tanaka, Yokoyama,
  Balatsky, and Nagaosa}}]{TYBN09}
\bibinfo{author}{\bibfnamefont{Y.}~\bibnamefont{Tanaka}},
  \bibinfo{author}{\bibfnamefont{T.}~\bibnamefont{Yokoyama}},
  \bibinfo{author}{\bibfnamefont{A.~V.} \bibnamefont{Balatsky}},
  \bibnamefont{and} \bibinfo{author}{\bibfnamefont{N.}~\bibnamefont{Nagaosa}},
  \bibinfo{journal}{Phys.\ Rev.\ B} \textbf{\bibinfo{volume}{79}},
  \bibinfo{pages}{06505(R)} (\bibinfo{year}{2009}).

\bibitem[{\citenamefont{Sato}()}]{Sato09}
\bibinfo{author}{\bibfnamefont{M.}~\bibnamefont{Sato}}, \eprint{in
  preparation}.

\bibitem[{\citenamefont{Read and Green}(2000)}]{RG00}
\bibinfo{author}{\bibfnamefont{N.}~\bibnamefont{Read}} \bibnamefont{and}
  \bibinfo{author}{\bibfnamefont{D.}~\bibnamefont{Green}},
  \bibinfo{journal}{Phys.\ Rev.\ B} \textbf{\bibinfo{volume}{61}},
  \bibinfo{pages}{10267} (\bibinfo{year}{2000}).

\end{thebibliography}

\end{document}